\documentstyle[12pt]{article}

\begin{document}
\setcounter{page}{1}
\pagestyle{plain} \vspace{1cm}
\begin{center}
\Large{\bf Gauss-Bonnet Braneworld Cosmology with Modified Induced Gravity on the Brane}\\
\small \vspace{1cm} {\bf Kourosh
Nozari$^{a,}$\footnote{knozari@umz.ac.ir}}, \quad {\bf Faeze
Kiani$^{b,}$\footnote{fkiani@umz.ac.ir}}\quad and \quad {\bf Narges
Rashidi$^{b,}$\footnote{n.rashidi@umz.ac.ir}}\\ \vspace{0.5cm} {
$^{a}$ Center for Excellence in Astronomy and Astrophysics
(CEAAI-RIAAM)-Maragha\\ P. O. Box: 55134-441, Maragha, IRAN}\\
\vspace{0.5cm} { $^{b}$Department of Physics, Faculty of Basic
Sciences, University of Mazandaran,\\
P. O. Box 47416-95447, Babolsar, IRAN}\\
\end{center}
\vspace{1.5cm}
\begin{abstract}
We analyze the background cosmology for an extension of the DGP
gravity with Gauss-Bonnet term in the bulk and $f(R)$ gravity on the
brane. We investigate implications of this setup on the late-time
cosmic history. Within a dynamical system approach, we study
cosmological dynamics of this setup focusing on the role played by
curvature effects. Finally we constraint the parameters of the model
by confrontation with recent observational data.
\end{abstract}
\vspace{1.5cm}
\newpage

\section{Introduction}
One of the most significant astronomical observation in the last
decade is the accelerated expansion of the universe [1-14]. One way
to explain this accelerating phase of the universe expansion is
invoking a dark energy component in the matter sector of the
Einstein field equations [15-31]. However, it is possible also to
modify the geometric part of the field equations to achieve this
goal [32-45]. In the spirit of the second viewpoint, braneworld
model proposed by Dvali, Gabadadze and Porrati (DGP) provides a
natural explanation of late-time accelerated expansion in its
self-accelerating branch of the solutions [46-51]. Unfortunately,
the self-accelerating branch of this scenario suffers from ghost
instabilities [52,53] and therefore it is desirable to invoke other
possibilities in this braneworld setup. An amazing feature of the
DGP setup is that the normal branch of this scenario, which is not
self-accelerating, has the capability to realize phantom-like
behavior without introducing any phantom field neither in the bulk
nor on the brane [54-60]. By the phantom-like behavior one means an
effective energy density which is positive, grows with time and its
equation of state parameter
($\omega_{eff}=\frac{p_{eff}}{\rho_{eff}}$) stays always less than
$-1$. The phantom-like prescription breaks down if this effective
energy density becomes negative. An interesting extension of the DGP
setup is possible modification of the induced gravity\footnote{We
call the $f(R)$ term on the brane the \emph{modified induced
gravity} since this braneworld scenario is an extension of the DGP
model. In the DGP model the gravity is induced on the brane through
interaction of the bulk graviton with loops of matter on the brane.
So, the phrase "Induced Gravity" is coming from the DGP character of
our model.} on the brane. This can be achieved by treating the
induced gravity in the framework of $f(R)$-gravity [61-66]. This
extension can be considered as a manifestation of the scalar-tensor
gravity on the brane since $f(R)$ gravity can be reconstructed as
General Relativity plus a scalar field [32-45]. Some features of
this extension such as self-acceleration in the normal branch of the
scenario are studied recently [61-68]. Here we generalize this
viewpoint to the case that the Gauss-Bonnet curvature effect is also
taken into account. We consider a DGP-inspired braneworld model that
the induced gravity on the brane is modified in the spirit of
$f(R)$-gravity and the bulk action contains the Gauss-Bonnet term to
incorporate higher order curvature effects. Our motivation is to
study possible influences of the curvature corrections on the
cosmological dynamics on the normal branch of the DGP setup. We
analyze the background cosmology and possible realization of the
phantom-like behavior in this setup. By introducing a
\emph{curvature fluid} that plays the role of dark energy component,
we show that this model realizes phantom-like behavior on the normal
branch of the scenario in some subspaces of the model parameter
space, without appealing to phantom fields neither in the bulk nor
on the brane and by respecting the null energy condition in the
phantom-like phase of expansion. We show also that in this setup
there is smooth crossing of the phantom divide line by the equation
of state parameter and the universe transits smoothly from
quintessence-like phase to a phantom-like phase. We present a
detailed analysis of cosmological dynamics in this setup within a
dynamical system approach in order to reveal some yet unknown
features of this kinds of models in their phase space. Finally
confrontation of our model with recent observational data leads us
to some constraint on model parameters.

\section{The Setup}
\subsection{Gauss-Bonnet Braneworld with Induced Gravity on the Brane}
The action of a GBIG (the Gauss-Bonnet term in the bulk and Induced
Gravity on the brane) braneworld scenario can be written as follows
[69-80]
\begin{equation}
S=S_{bulk}+S_{brane}
\end{equation}
where by definition
\begin{equation}
S_{bulk}=\frac{1}{16\pi G^{(5)}}\int d^{5}x\sqrt{-g}[{\cal
R}^{(5)}-2\Lambda^{(5)}+\alpha {\cal{L}}_{GB}],
\end{equation}
with
\begin{equation}
{\cal{L}}_{GB}=\big({\cal R}^{(5)}\big)^{2}-4{\cal
R}^{(5)}_{AB}{\cal R}^{(5)AB}+{\cal R}^{(5)}_{ABCD}{\cal
R}^{(5)ABCD},
\end{equation}
and
\begin{equation}
S_{brane}=\frac{1}{16\pi G^{(4)}}\int d^{4}x\sqrt{-q}\Big[{\cal
R}-2\Lambda^{(4)}\Big].
\end{equation}
$G^{(5)}$ is the 5D Newton's constant in the bulk and $G^{(4)}$ is
the corresponding 4D quantity on the brane. ${\cal{L}}_{GB}$ is the
Gauss-Bonnet term, and $\alpha$ is a constant with dimension of
$[length]^{2}$. $q$ is the induced metric on the brane. We choose
the coordinate of the extra dimension to be $y$, so that the brane
is located at $y=0$. The DGP crossover distance which is defined as
\begin{equation}
r_{c}=\frac{G^{(5)}}{G^{(4)}}=\frac{\kappa_{5}^{2}}{2\kappa_{4}^{2}},
\end{equation}
has the dimension of $[length]$ and will be appeared in our
forthcoming equations. We note that this scenario is UV/IR complete
in some sense, since it contains both the Gauss-Bonnet term as a
string-inspired modification of the UV (ultra-violet) sector and the
induced gravity as IR (infra-red) modification to the General
Relativity. The cosmological dynamics of this GBIG scenario is
described fully by the following Friedmann equation [81-84]
\begin{equation}
\bigg[H^{2}+\frac{k}{a^{2}}-\frac{8\pi
G^{(4)}(\rho+\lambda)}{3}\bigg]^{2}=
\frac{4}{r_{c}^{2}}{\bigg[1+\frac{8\alpha}{3}\Big(H^{2}+\frac{k}{a^{2}}+\frac{U}{2}\Big)\bigg]}^{2}
\Big(H^{2}+\frac{k}{a^{2}}-U\Big)\,,
\end{equation}
where
\begin{equation}
U=-\frac{1}{4\alpha}\pm
\frac{1}{4\alpha}\sqrt{1+4\alpha\Big(\frac{\Lambda^{(5)}}{6}+\frac{2{\cal{E}}_{0}G^{(5)}}{a^{4}}\Big)}\,,
\quad\quad \lambda\equiv\frac{\Lambda^{(4)}}{8\pi G^{(4)}}.
\end{equation}
${\cal{E}}_{0}$ is referred hypothetically as the mass of the bulk
black hole and the corresponding term is called the bulk radiation
term. Note that when one adopts the positive sign, the above
equation can be reduced to the generalized DGP model as
$\alpha\rightarrow0$\,, but the branch with negative sign cannot be
reduced to the generalized DGP model in this regime. Therefore, we
just consider the plus sign of the above equation [81-84]. We note
that depending on the choice of the bulk space, the brane FRW
equations are different (see [85] for details). The bulk space in
the present model is a 5-dimensional AdS black hole. In which
follows, we assume that there is no cosmological constant on the
brane and also in the bulk, \emph{i.e.}
$\Lambda^{(4)}=\Lambda^{(5)}=0$. Also we ignore the bulk radiation
term since its decay very fast in the early stages of the evolution
(note however that this term is important when one treats
cosmological perturbations on the brane). So, the Friedmann equation
in this case reduces to the following form
\begin{equation}
\bigg[H^{2}-\frac{8\pi
G^{(4)}}{3}\rho\bigg]^{2}=\frac{4}{r_{c}^{2}}{\bigg[1+\frac{8\alpha}{3}H^{2}\bigg]}^{2}
H^{2}.
\end{equation}
It has been shown that it is possible to realize the phantom-like
behavior in this setup without introducing any phantom matter on the
brane [86-91]. In which follows we generalize this setup to the case
that induced gravity on the brane is modified in the spirit of
$f(R)$ gravity and we explore the cosmological dynamics of this
extended braneworld scenario.

\subsection{Modified GBIG Gravity}
In this subsection we firstly formulate a GBIG scenario that induced
gravity on the brane acquires a modification in the spirit of $f(R)$
gravity. To obtain the generalized Friedmann equation of this model
we proceed as follows: firstly, the Friedmann equation for pure DGP
scenario is as follows [49-51,92,93]
\begin{equation}
\epsilon\sqrt{H^{2}-\frac{{\cal{E}}_{0}}{a^{4}}-\frac{\Lambda_{5}}{6}+
\frac{k}{a^{2}}}=r_{c}\bigg[(H^{2}+\frac{k}{a^{2}})-\frac{8\pi
G^{(4)}}{3}(\rho+\lambda)\bigg]
\end{equation}
where $\epsilon=\pm1$ is corresponding to two possible embeddings of
the brane in the bulk. Considering a Minkowski bulk with
$\Lambda_{5}=0$ and by setting ${\cal{E}}_{0}=0$ with a tensionless
brane ($\lambda=0$), for a flat brane $(k=0)$ we find
\begin{equation}
H^{2}=\frac{8\pi G^{(4)}}{3}\rho\pm\frac{H}{r_{c}}.
\end{equation}
The normal branch of the scenario is corresponding to the minus sign
in the right hand side of this equation. The second term in the
right is the source of the phantom-like behavior on the normal
branch: the key feature of this phase is that the brane is
extrinsically curved in such a way that shortcuts through the bulk
allow gravity to screen the effects of the brane energy-momentum
contents at Hubble parameters $H\sim r_{c}^{-1}$ and this is not the
case for the self-accelerating phase [54-60].

In the next step, we incorporate possible modification of the
induced gravity by inclusion of a $f(R)$ term on the brane. This
extension can be considered as a manifestation of the scalar-tensor
gravity on the brane. In this case we find the following generalized
Friedmann equation (see for instance [61-68,92,93])

$$\epsilon\sqrt{H^{2}-\frac{M G^{(5)}}{a^{4}}-\frac{\Lambda_{5}}{6}+
\frac{k}{a^{2}}}=$$
\begin{equation}
r_{c}\bigg[\Big(H^{2}+\frac{k}{a^{2}}\Big)f'(R)-\frac{8\pi
G^{(4)}}{3}\Big[\rho+\lambda+\Big(\frac{1}{2}\big[Rf'(R)-f'(R)\big]-3H\dot{R}f''(R)\Big)\Big]\bigg]\,,
\end{equation}
where a prime marks derivative with respect to $R$. In the third
step, we need to the GBIG Friedmann equation in the absence of any
modification of the induced gravity on the brane, that is, without
$f(R)$ term on the brane. This has been obtained in the previous
subsection, equation (6). Now we have all prerequisite to obtain the
Friedmann equation of our GBIG-modified gravity scenario. The
comparison between previous equations gives this Friedmann equation
of cosmological dynamics as follows\footnote{We note that this
equation can be derived using the generalized junction conditions on
the brane straightforwardly, see [81-84].}
\begin{equation}
H^{2}=\frac{\kappa_{4}^{2}}{3f'(R)}\rho+\frac{\kappa_{4}^{2}}{3}\rho_{curv}
\pm \frac{1}{r_{c}f'(R)}\Big(1+\frac{8}{3}\alpha H^{2}\Big)H.
\end{equation}
where we have defined hypothetically the following energy density
corresponding to curvature effect
\begin{equation}
\rho_{curv}=\frac{1}{f'(R)}
\Big(\frac{1}{2}[Rf'(R)-f(R)]-3H\dot{R}f''(R)\Big).
\end{equation}
Note that to obtain this relation,  ${\cal{E}}_{0}$, $\Lambda_{5}$,
$\lambda$ and $k$\, are set equal to zero. From now on we restrict
our attention to the normal branch of the scenario, \emph{i.e.} the
minus sign in equation (12) because there is no ghost instabilities
in this branch if only the DGP character of the model is considered.

Note however that although we refer to the normal (ghost-free)
branch of the DGP model (in the sense that for $f(R)= R$ the
obtained solutions reduce to this branch) as an indication of the
ghost-free property of the considered solutions, it is not a-priori
guaranteed that on the obtained dS backgrounds which generalize the
normal DGP branch, the ghost does not reappear. In fact, the ghost
on the self-accelerated branch of the DGP model is entirely the
problem of the de Sitter background. Within the crossover scale
$r_c$ which is the horizon for the self-accelerated branch, the
theory reduces to a scalar-tensor model, with the scalar sector
(brane bending mode) described by a simple Galileon self-interaction
[94] as $${\cal{L}}_{\pi}=\pi\Box \pi -\frac{(\partial \pi)^{2}\Box
\pi}{\Lambda^{3}},$$ which, in spite of the presence of higher
derivatives, propagates a single healthy degree of freedom. The
ghost on the self-accelerated branch arises merely due to the fact
that $\pi$ gets a nontrivial profile and the kinetic term for its
perturbation flips the sign on the background. A similar argument
can be applied in the present work to overcome the ghost
instabilities in this extended braneworld setup.

In which follows, we assume that the energy density $\rho$ on the
brane is due to cold dark matter (CDM) with
$\rho_{m}=\rho_{0m}(1+z)^{3}$. We can rewrite the Friedmann equation
in terms of observational parameters such as the redshift $z$ and
dimensionless energy densities $\Omega_{i}$ as follows
\begin{equation}
E^{2}=\frac{\Omega_{m}}{f'(R)}(1+z)^{3}+
\Omega_{curv}(1+z)^{3(1+\omega_{curv})}
-2\frac{\sqrt{\Omega_{r_{c}}}}{f'(R)}\Big[1+\Omega_{\alpha}E^{2}(z)\Big]E(z)
\end{equation}
where$$E(z)\equiv\frac{H}{H_{0}}$$
$$\Omega_{m}\equiv\frac{\kappa_{4}^{2}}{3H_{0}^{2}}\rho_{0m},\quad\Omega_{\alpha}\equiv\frac{8}{3}\alpha
H_{0}^{2},\quad
\Omega_{r_{c}}\equiv\frac{1}{4r_{c}^{2}H_{0}^{2}},\quad
\Omega_{curv}\equiv\frac{\kappa_{4}^{2}}{3H^{2}_{0}}\rho_{0curv}$$
and
$$\omega_{curv}=\frac{p_{curv}}{\rho_{curv}}.$$ $p_{curv}$, the hypothetical pressure of
the curvature effect, can be obtained by the following equation of
continuity
\begin{equation}
\dot{\rho}_{curv}+3H\bigg(\rho_{curv}+p_{curv}+\frac{\dot{R}f''(R)}{r_{c}[f'(R)]^{2}}\bigg)=\frac{3H_{0}^{2}\Omega_{m}\dot{R}f''(R)}{[f'(R)]^{2}}
a^{-3}.
\end{equation}
One can obtain a constraint on the cosmological parameters of the
model at $z=0$ as follows
\begin{equation}
\Omega_{m}=1-\Omega_{curv}+2\sqrt{\Omega_{r_{c}}}(1+\Omega_{\alpha}).
\end{equation}
Note that we have used the normalization $f'(R)|_{z=0}=1$ in this
relation which is observationally a viable assumption.

\section{Cosmological dynamics in the modified GBIG scenario}

Now we study cosmological dynamics in this setup. To solve the
Friedmann equation for the normal branch of this scenario, it is
convenient (following the papers by Bohamdi-Lopez in Ref. [86-90])
to introduce the dimensionless variables as follows
\begin{equation}
\bar{H}\equiv\frac{8}{3}\frac{\alpha}{r_{c}f'(R)}H=2\Omega_{\alpha}\sqrt{\Omega_{r}}E(z)
\end{equation}
\begin{equation}
\bar{\rho}\equiv\frac{32}{27}\frac{\kappa_{5}^{2}\alpha^{2}}{[r_{c}f'(R)]^3}\Big(\rho_{m}+
f'(R) \rho_{curv}\Big)
=4\Omega_{\alpha}^2\Omega_{r}\Big[\Omega_{m}(1+z)^3+\Omega_{curv}(1+z)^{3(1+\omega_{curv})}\Big]
\end{equation}
\begin{equation}
b\equiv\frac{8}{3}\frac{\alpha}{[r_{c}f'(R)]^2}=4\Omega_{\alpha}\Omega_{r}.
\end{equation}
An effective crossover distance which is appeared on the right hand
side of these relations can be defined as follows
\begin{equation}
r\equiv r_{c}f'(R),
\end{equation}
and by definition $\Omega_{r}\equiv\frac{1}{4r^{2}H_{0}^{2}}$. Then
the Friedmann equation can be rewritten as
\begin{equation}
\bar{H}^3+\bar{H}^2+b\bar{H}-\bar{\rho}=0.
\end{equation}
The number of real roots of this equation is determined by the sign
of the discriminant function ${\cal{N}}$ defined as
\begin{equation}
{\cal{N}}=Q^3+S^2
\end{equation}
where $Q$ and $S$ are defined as
$$Q=\frac{1}{3}\Big(b-\frac{1}{3}\Big)$$ and
$$S=\frac{1}{6}b+\frac{1}{2}\bar{\rho}-\frac{1}{27}$$
respectively. We can rewrite ${\cal{N}}$ as
\begin{equation}
{\cal{N}}=\frac{1}{4}\Big(\bar{\rho}-\bar{\rho}_{1}\Big)\Big(\bar{\rho}-\bar{\rho}_{2}\Big),\\
\end{equation}
where
\begin{equation}
\bar{\rho}_{1}=-\frac{1}{3}\Big\{b-\frac{2}{9}[1+\sqrt{(1-3b)^3}\,]\Big\}
\end{equation}
\begin{equation}
\bar{\rho}_{2}=-\frac{1}{3}\Big\{b-\frac{2}{9}[1-\sqrt{(1-3b)^3}\,]\Big\}.
\end{equation}
In which follows, we consider just the real and positive roots of
the Friedmann equation (21). For $0<b<\frac{1}{4}$,\,
$\bar{\rho}_{1}>0$ and $\bar{\rho}_{2}<0$. Then, the number of real
roots of the cubic Friedmann equation depends on the minimum energy
density of the brane and the situation of $\rho_{1}$ relative to
this minimum. Since in our setup, curvature effect plays the role of
the dark energy component on the brane, we can consider two
different regimes to determine the minimum value of $\bar{\rho}$
as follows: \\

1.\quad{$\omega_{curv}>-1$}\\

In this case, curvature fluid plays the role of quintessence
component, then the minimum value happens asymptotically at $z=-1$
and we will obtain $\bar{\rho}_{min}=0$. In this situation we can
define three possible regimes: a high energy regime with
$\bar{\rho}_{1} < \bar{\rho}$\,;\, a limiting regime with
$\bar{\rho}_{1}=\bar{\rho}$ \, and  a low energy regime with
$\bar{\rho}_{1}> \bar{\rho}$. In each of these cases, depending on
the sign of $N$ there are different solutions [86-91].\\

2.\quad{$\omega_{curv}<-1$}\\

In this case, the curvature fluid plays the role of a phantom
component (we will investigate its phantom-like behavior in the next
section) and the minimum value of $\bar{\rho}$ happens at $z=0.18$.
So we find the $\bar{\rho}_{min}$ as follows
\begin{equation}
\bar{\rho}_{min}=4\Omega_{\alpha}^{2}\Omega_{r_{c}}\Big[0.43+\Omega_{curv}(1+0.18)^{3(1+\omega_{curv})}\Big]
\end{equation}
where we have set $\Omega_{m}=0.26$. We note that $w_{curv}$ is not
constant and as we will show, it evolves from quintessence to
phantom phase. We note also that the value of redshift that
$\bar{\rho}_{min}$ occurs (that is, $z=0.18$), has no dependence on
the values of $w_{curv}$. Here we treat only the case
$\bar{\rho}_{1}<\bar{\rho}_{min}$ with details. When this condition
is satisfied, the function ${\cal{N}}$ is positive and there is a
unique solution for expansion of the brane described by
\begin{equation}
\bar{H}_{1}=\frac{1}{3}\Big[2\sqrt{1-3b}\cosh(\frac{\eta}{3})\Big]
\end{equation}
where $\eta$ is defined as
\begin{equation}
\cosh(\eta)=\frac{S}{\sqrt{-Q^3}}
\end{equation}
We note that this condition provides a constraint on the
dimensionless parameters $\Omega_{i}$ as follows
\begin{equation}
-\frac{1}{3b}\Big[b-\frac{2}{9}[1+\sqrt{(1-3b)^3}]\Big]<\Omega_{\alpha}
\Big[0.43+\Omega_{curv}(1+0.18)^{3(1+\omega_{curv})}\Big]
\end{equation}
\begin{figure}[htp]
\begin{center}\includegraphics{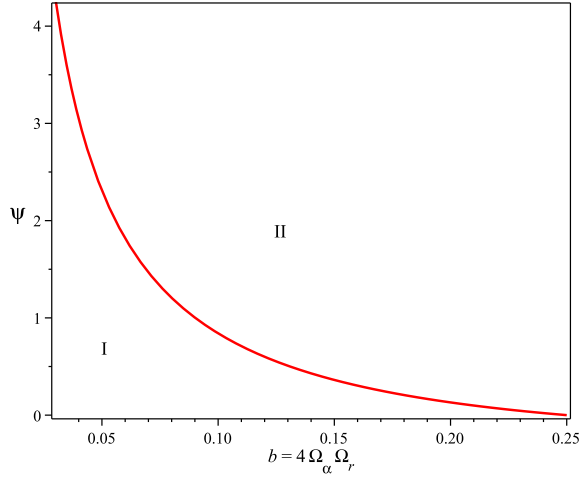} \vspace{8cm}
\end{center}
\caption{\small {The upper region (region II) of this figure is
corresponding to the set $( \Omega_{r},\, \Omega_{\alpha},\,
\Omega_{curv},\, \omega_{curv} )$ that fulfil the inequality (29).}}
\end{figure}
Figure $1$ shows the phase space of the above relation. In this
figure we have defined $\Psi\equiv
0.43+\Omega_{curv}(1+0.18)^{3(1+\omega_{curv})}$. The relation (29)
is fulfilled for upper region (region II) of this figure. In this
case there are three possible regimes as was mentioned above. A
point that should be emphasized here is the fact that in the
presence of modified induced gravity on the brane, the solution of
the generalized Friedmann equation (12) is actually rather involved
due to simultaneously presence of $H$, $\dot{H}$ and $\ddot{H}$. A
through analysis of this problem is out of the scope of this study,
but there are some attempts (such as cosmography) in this direction
to construct an operational framework to treat this problem, see for
instance [95]. Here we have tried to find a solution of equation
(12) by using the discriminant function ${\cal{N}}$, the result of
which is given by (27). However, we note that a complete analysis is
needed for instance in the framework of cosmography of brane $f(R)$
gravity [96].

To investigate cosmology described by solution (27), we rewrite the
original Friedmann equation in the following form in order to create
a general relativistic description of our model
\begin{equation}
H^{2}=\frac{(\kappa_{4}^{2})_{eff}}{3}\rho_{m}+\rho_{curv}-\frac{H}{r_{c}f'(R)}\Big(1+\frac{8}{3}\alpha
H^{2}\Big)
\end{equation}
that $ \rho_{curv}$ is defined in (13). Comparing this relation with
the Friedmann equation in GR
\begin{equation}
H^{2}=\frac{\kappa_{4}^{2}}{3}(\rho_{m}+\rho_{eff}),
\end{equation}
we obtain an effective energy density
\begin{equation}
\frac{\kappa_{4}^{2}}{3}\rho_{eff}=\rho_{curv}-\frac{H}{r_{c}f'(R)}\Big(1+\frac{8}{3}\alpha
H^{2}\Big),
\end{equation}
which can be rewritten as follows  \begin{equation}
\rho_{eff}=\Omega_{curv}(1+z)^{3(1+\omega_{curv})}-\frac{2\sqrt{\Omega_{r_{c}}}}{f'(R)}(1+\Omega_{\alpha}
E^{2})E(z).
\end{equation}
The dependence of $\rho_{eff}$ on the redshift depends itself on the
regimes introduced above and the form of $f(R)$ function. Figure $2$
shows the variation of $\rho_{eff}$ versus the redshift for
\begin{equation}
f(R)=R-(n-1)\zeta^{2}\Big(\frac{R}{\zeta^{2}}\Big)^{n}
\end{equation}
with $n=0.25$. This value of $n$ lies well in the range of
observationally acceptable values of $n$ from Solar System
tests\footnote{We note however that the key issue with regards to
passing Solar-System tests is not the value of $n$, but the value of
$f'(R)$ today. In fact experimental data tell us that $f'(R)-1 ~ <
10^{-6}$, when $f'(R)$ is parameterized to be exactly $1$ in the far
past.} [32-45].

\begin{figure}[htp]
\begin{center}\includegraphics{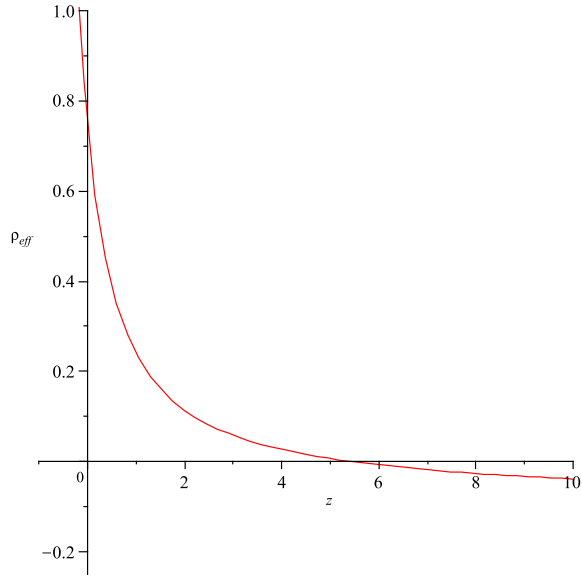} \vspace{7cm}
\end{center}
\caption{\small {Variation of effective energy density versus the
redshift for an specific $f(R)$ model described as (34) with
$n=0.25$.}}
\end{figure}
The effective energy density shows a phantom-like behavior
\emph{i.e.} it increases with cosmic time. This is a necessary
condition to have phantom-like behavior but it is not sufficient: we
should check status of the deceleration parameter and also equation
of state parameter. In a general relativistic description of our
model, one can rewrite the energy conservation equation as follows
\begin{equation}
\dot{\rho}_{eff}+3H(1+\omega_{eff})\rho_{eff}=0
\end{equation}
which leads to the following relation for $1+\omega_{eff}$
$$1+\omega_{eff}=-\frac{\dot{\rho}_{eff}}{3H\rho_{eff}}$$
Figure $3$ shows variation of the effective equation of state
parameter versus the redshift for an specific $f(R)$ model described
as (34) with $n=0.25$. The effective equation of state parameter
transits to the phantom phase $1+\omega_{eff}<0$\,  but there is no
smooth crossing of the phantom divide line in this setup. We note
that adopting other general ansatz such as the Hu-Sawicki model [97]
\begin{eqnarray}
f(R)=R-R_{c}\frac{\alpha_{0}(\frac{R}{R_{c}})^{n}}{1+\beta_{0}(\frac{R}{R_{c}})^{n}}
\end{eqnarray}
(where both $\alpha$ and $R_{c}$ are free positive parameters), does
not change this result in our framework.
\begin{figure}[htp]
\begin{center}\includegraphics{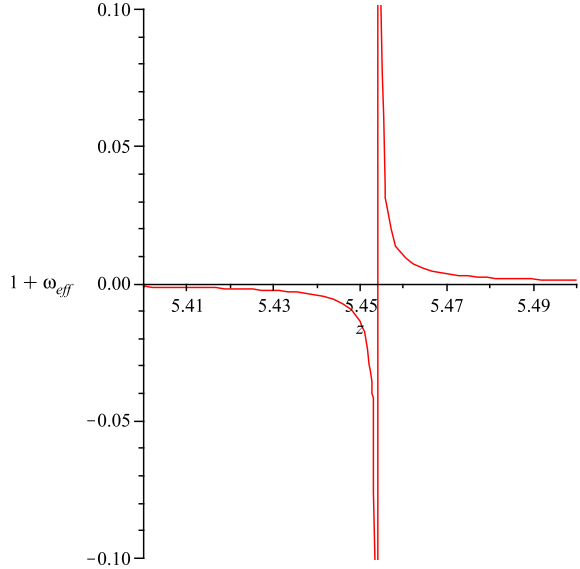} \vspace{7.5cm}
\end{center}
\caption{\small {Variation of effective equation of state parameter
versus the redshift for an specific $f(R)$ model described as (34)
with $n=0.25$.}}
\end{figure}
The deceleration parameter defined as
$$q\equiv-1-\frac{\dot{H}}{H^{2}}$$ takes the following form in our
setup
$$q=-1-\frac{\Omega_{m}(1+z)^{3}\Big(\frac{3}{2}-\frac{\dot{f}'(R)}{f'(R)}\frac{H_{0}}{E(z)}\Big)}
{f'(R)E^{2}(z)+\sqrt{\Omega_{r_{c}}}\Big(1+3\Omega_{\alpha}E^{2}(z)\Big)E(z)}$$
\begin{equation}
-\frac{\sqrt{\Omega_{r_{c}}}\frac{\dot{f}'(R)}{f'(R)}(1+\Omega_{\alpha}E^{2}(z))
\frac{1}{H_{0}}-\frac{3}{2}\Omega_{curv}f'(R)(1+z)^{3(1+w_{c})}(1+w_{c})}
{f'(R)E^{2}(z)+\sqrt{\Omega_{r_{c}}}\Big(1+3\Omega_{\alpha}E^{2}(z)\Big)E(z)}.
\end{equation}
In this relation $H$ and $f(R)$ are defined as (27) and (34). Figure
$4$ show variation of $q$ versus the redshift for $n=0.25$.
\begin{figure}[htp]
\begin{center}\includegraphics{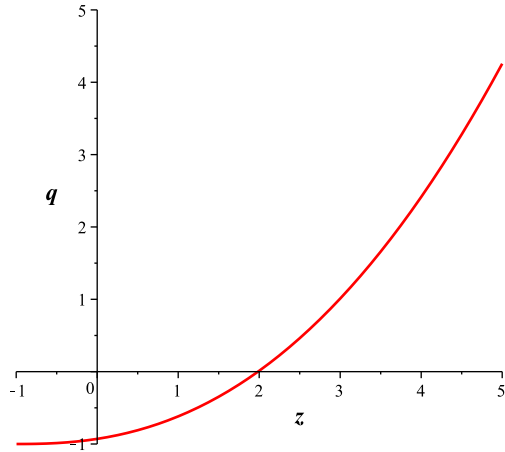} \vspace{7cm}
\end{center}
\caption{\small {Variation of the deceleration parameter $q$ versus
$z$ for $n=0.25$. The transition from deceleration to the
acceleration phase occurs at $z\approx 2$. }}
\end{figure}
The universe enters the accelerated phase of expansion at
$z\simeq2$. Another important issue to be investigated in this setup
is the big-rip singularity. To avoid super-acceleration on the
brane, it is necessary to show that Hubble rate decreases as the
brane expands and there is no big-rip singularity in the future.
Figure $5$ shows variation of $\dot{H}$ versus $z$. We see that in
this model $\dot{H}<0$ always and therefore, there is no
super-acceleration and future big-rip singularity in this setup.
\begin{figure}[htp]
\begin{center}\includegraphics{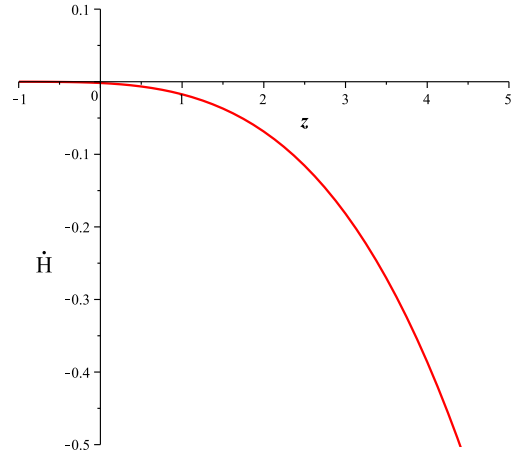} \vspace{6cm}
\end{center}
\caption{\small {In this model $\dot{H}$ is always negative and
therefore there is no super-acceleration and big-rip singularity. }}
\end{figure}
All the previous considerations show that this model accounts for
realization of the phantom-like behavior without introducing a
phantom field neither on the brane nor in the bulk. Nevertheless, we
have to check the status of the null energy condition in this setup.
Figure $6$ shows the variation of $(\rho+p)_{tot}$ versus the
redshift. We see that this condition is fulfilled at least in some
subspaces of the phantom-like region of the model parameter space.

\begin{figure}[htp]
\begin{center}\includegraphics{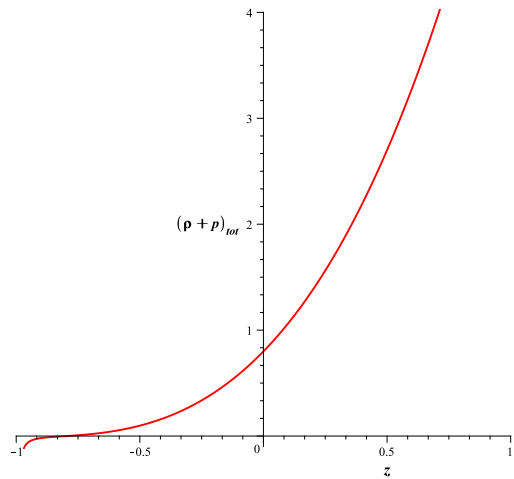} \vspace{6cm}
\end{center}
\caption{\small {The status of the null energy condition in this
model. }}
\end{figure}
\newpage

\section{A dynamical system viewpoint}
Up to this point, we have shown that there are effective quantities
that create an effective phantom-like behavior on the brane. In this
respect, one can define a potential related to the effective phantom
scalar field $\phi_{eff}$ as follows [98]
\begin{equation}
\frac{V_{eff}(z)}{3H_{0}^{2}}=E^{2}-\frac{\Omega_{m}(1+z)^{3}}{f'(R)}+\frac{1}{2}\frac{d(E^{2}-
\frac{\Omega_{m}(1+z)^{3}}{f'(R)})}{d\ln{(1+z)^{3}}}
\end{equation}
\begin{equation}
\frac{\phi_{eff}(z)}{\sqrt{3}}=-\int
\frac{dz}{(1+z)E}\sqrt{\frac{d(E^{2}-
\frac{\Omega_{m}(1+z)^{3}}{f'(R)})}{d\ln{(1+z)^{3}}}}\,.
\end{equation}
We note that in principle these equations can lead to
$V_{eff}=V_{eff}(\phi_{eff})$, but in practice the inversion cannot
be performed analytically. Now we define the following normalized
expansion variables [99-102]
\begin{equation}
x=\frac{\sqrt{\Omega_{m}}}{a^{\frac{3}{2}}E}\,,\quad
y=\frac{\sqrt{\Omega_{curv}f'(R)}}{a^{\frac{3}{2}(1+w_{curv})}E}\,,\quad
\upsilon=\frac{\sqrt{\Omega_{r_{c}}}}{E}\,,\quad
\chi=\sqrt{\Omega_{\alpha}}E.
\end{equation}
With these definitions, the Friedmann equation (14) takes the
following form
\begin{equation}
x^{2}+y^{2}-2\upsilon(1+\chi^{2})=f'(R)
\end{equation}
This constrain means that the phase space can be defined by the
relation $x^{2}+y^{2}-2\upsilon\geq1$, since $f'(R)-1<10^{-6}$ by
solar system constraints and $\upsilon$ is a positive quantity.
Introducing the new time variable $\tau=\ln{a}$, and eliminating
$\chi$ and $E$, we obtain the following autonomous system
\begin{equation}
x'=x(q-\frac{1}{2})\,,
\end{equation}
\begin{equation}
y'=\frac{y}{y^{2}-1}\bigg[\frac{3}{2}(1+w_{curv})(1+x^{2})-(q+1)\Big[1+x^{2}-2v+\frac{1}{2}(2+v)(x^{2}+y^{2}-1)\Big]\bigg]\,,
\end{equation}
\begin{equation}
\upsilon'=\upsilon(q+1)\,.
\end{equation}
Here primes denote differentiation with respect to $\tau$, and
$q=-\frac{\ddot{a}a}{\dot{a}^{2}}$ stands for the deceleration
parameter
\begin{equation}
q+1=\frac{\frac{3}{2}\Big(x^{2}+[1+w_{curv}(1-x^{2})]y^{2}\Big)}{4x^{2}+8y^{2}-x^{2}y^{2}-
\frac{3}{2}y^{4}+\frac{1}{4}\upsilon y^{4}-\frac{9}{4}\upsilon
y^{2}-\frac{1}{2}\upsilon x^{2}-\frac{19}{2}\upsilon-5}.
\end{equation}
To study cosmological evolution in the dynamical system approach, it
is necessary to find fixed (or critical) points of the model that
are denoted by $(x^{*}, y^{*}, \upsilon^{*})$. These points are
achieved by fulfillment of the following condition
\begin{equation}
g^{i}(x^{*}, y^{*}, \upsilon^{*})= 0
\end{equation}
where
\begin{equation}
{x'}^{i}=g^{i}(x^{*}, y^{*}, \upsilon^{*})
\end{equation}
A part of dynamical system analysis of this model is summarized in
table $1$.
\begin{table}
\begin{center}
\caption{Location and deceleration parameter of the critical points.
The location of point $B$ and the deceleration factor of points $C$
and $D$ are dependent on the equation of state parameter of the
perfect fluid, $w_{curv}$\,. The critical curve $F$ exists just for
$w_{curv}=-1$ } \vspace{1 cm}
\begin{tabular}{|c|c|c|c|c|c|c|}
  \hline
  \hline $Name$ &$x$&$y$&$\upsilon$&$q$\\
  \hline
     \hline $A$ &$\frac{\sqrt{15}}{3}$&$0$&$0$ & $\frac{1}{2}$ \\
     \hline
     \hline $B$&$x_{_B}$&$y_{_B}$&$0$&$\frac{1}{2}$ \\
   \hline
   \hline $C$&$0$&$\frac{1}{2}$&$0$&$-(1.17+0.17w_{curv})$ \\
   \hline
   \hline $D$&$0$&$\frac{7}{2}$&$0$&$-(1.13+0.13w_{curv})$ \\
   \hline
   \hline $E$&$0$&$0$&$v^{*}$&$-1$ \\
   \hline
   \hline $F$&$0$&$y^{*}$&$v^{*}$&$-1$ \\
   \hline
\end{tabular}
\end{center}
$$$$
where in this table $x_{_B}$ and $y_{_B}$ are as follows
$$y_{_B}=\sqrt {x_{_B}^{2}(w_{curv}-1)+(1+w_{curv})}$$ and
$$x_{_B}=\sqrt {{\frac {-17\,w_{curv}+4+6\,w^{2}_{curv}+\sqrt {297\,w^{2}_{curv}-120\,w_{curv}-180\,
w_{curv}^{3}+16+36\,w_{curv}^{4}}}{2\,(w_{curv}-1)}}}
$$
\end{table}

The critical points $A$ and $B$ demonstrate the early-time, matter
dominated epoch which lead to a positive deceleration parameter.
Points $C$ and $D$ which are phases with vanishing matter character,
that is $\Omega_{m}=0$, can explain the positively accelerated phase
of the universe expansion for $\omega_{curv}\geq -6.88$. Critical
curve $E$ also demonstrates a positively accelerated phase for all
values of the equation of state parameter of the curvature fluid.
Critical curve $F$ which exists only for the case
$\omega_{curv}=-1$, is a de Sitter phase in this model. Existence of
a stable de Sitter point and an unstable matter dominated phase (in
addition to radiation dominated era) in the universe expansion
history is required for cosmological viability of any cosmological
model. In order to investigate the stability of these points, one
can obtain the eigenvalues of these points separately. Based on
table 2, in order for point $A$ to be an unstable point, it is
necessary to have $\omega_{curv}<\frac{1}{4}$\,. Therefore, the
point $A$ as a saddle point agrees with what we have shown in figure
$7$. Now the stability of the positively accelerated phases of the
model depends on whether the curvature fluid plays the role of a
quintessence scalar field or not. Points $C$, $D$ and $E$ of table 2
are stable phases of this model if $\omega_{curv}>-1$. Whereas if
the curvature fluid plays the role of a cosmological constant, the
point $F$ will be a stable de Sitter phase. We note that generally,
if a nonlinear system has a critical curve, the Jacobian matrix of
the linearized system at a critical point on the curve (line in our
2-dimensional subspace) has a zero eigenvalue with an associated
eigenvector tangent to the critical curve at the chosen point. The
stability of an specific critical point on the curve can be
determined by the nonzero eigenvalues, because near this critical
point there is essentially no dynamics along the critical curve
[103]. We have plotted the phase space of this model in subspace
$x-y$ with $\omega_{curv}=-1$. As we see, point $A$ is a saddle
point and curve $F$ is a stable de Sitter curve.

\begin{table}
\begin{center}
\caption{Eigenvalues and the stability regions of the critical
points. $\varsigma(w_{curv})$, $\xi(w_{curv})$ and $\zeta
(w_{curv})$ are complicated functions of $w_{curv}$\,. The critical
curve $F$ for which $w_{curv}=-1$ is a stable de Sitter phase.}
\vspace{1 cm}
\begin{tabular}{|c|c|c|c|c|c|c|}
  \hline $Name$ & Eigenvalue& Stable region \\
  \hline
     \hline $A$ &$-9\,\,,-\frac{1}{2}\,\,,1-4w_{curv}$& $w_{curv}>\frac{1}{4}$ \\
     \hline
     \hline $B$&$\varsigma(w_{curv})\,,\xi(w_{curv})\,, \zeta (w_{curv})$&$$ \\
   \hline
   \hline $C$&$-(1+0.1w_{curv})\,,-(1+w_{curv})\,,-(2+0.17w_{curv})$& $w_{curv}>-1$ \\
   \hline
   \hline $D$&$-(1+0.1w_{curv})\,,-(1+w_{curv})\,,-(2+0.13w_{curv})$& $w_{curv}>-1$\\
   \hline
   \hline $E$&$-\frac{3}{2}\,\,,-\frac{3}{2}(1+w_{curv})\,\,,-2$&$w_{curv}>-1$ \\
   \hline
   \hline $F$&$-\frac{3}{2}\,\,,0\,\,,-2$& $w_{curv}=-1$ \\
   \hline

\end{tabular}
\end{center}
\end{table}
\newpage
\begin{figure}[htp]
\begin{center}\includegraphics{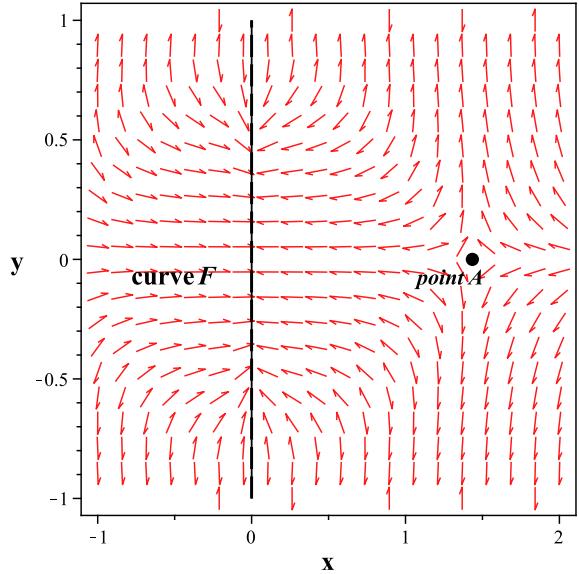} \vspace{6cm}
\end{center}
\caption{\small {The phase subspace $x-y$ that the curvature fluid
plays the role of a cosmological constant. This model is
cosmologically viable since there are an unstable matter dominated
point and a stable de Sitter phase. }}
\end{figure}

\section{Confrontation with Recent Observational Data}

In this section we use the combined data from Planck + WMAP + high
L+ lensing + BAO [104] to confront our model with recent
observation. In this way we obtain some constraints on the model
parameters; especially the Gauss-Bonnet curvature contribution. For
this purpose, we consider the relation between $\Omega_{curv}$ and
$\Omega_{m}$ in the background of the mentioned observational data.
We suppose that $\Omega_{curv}$ plays the role of dark energy in
this setup. Figure 8 shows the result of our numerical study. In
this model with $\Omega_{r_{c}}\sim 10^{-4}$\,(see [105] for
instance), $\Omega_{\alpha}$ is constraint as follows

$$\omega_{curv}=-0.5: \quad   0.008<\Omega_{\alpha}< 0.011$$
$$\omega_{curv}=-0.92: \quad   0.01<\Omega_{\alpha}< 0.055$$
$$\omega_{curv}=-1.05: \quad   0.012<\Omega_{\alpha}< 0.073$$

On the other hand, if we consider the $\Omega_{eff}$ defined as
$\Omega_{eff}=\frac{\kappa_{4}^{2}}{3H_{0}^{2}}\rho_{eff}$ as our
main parameter, the result will be as shown in figure 9. In this
case we have the following constraint on $\Omega_{\alpha}$

$$\Omega_{curv}=0.7: \quad  0.0043<\Omega_{\alpha}<0.036$$
$$\Omega_{curv}=0.5:  \quad  0.0048<\Omega_{\alpha}<0.0081$$

\begin{figure}[htp]
\begin{center}\includegraphics{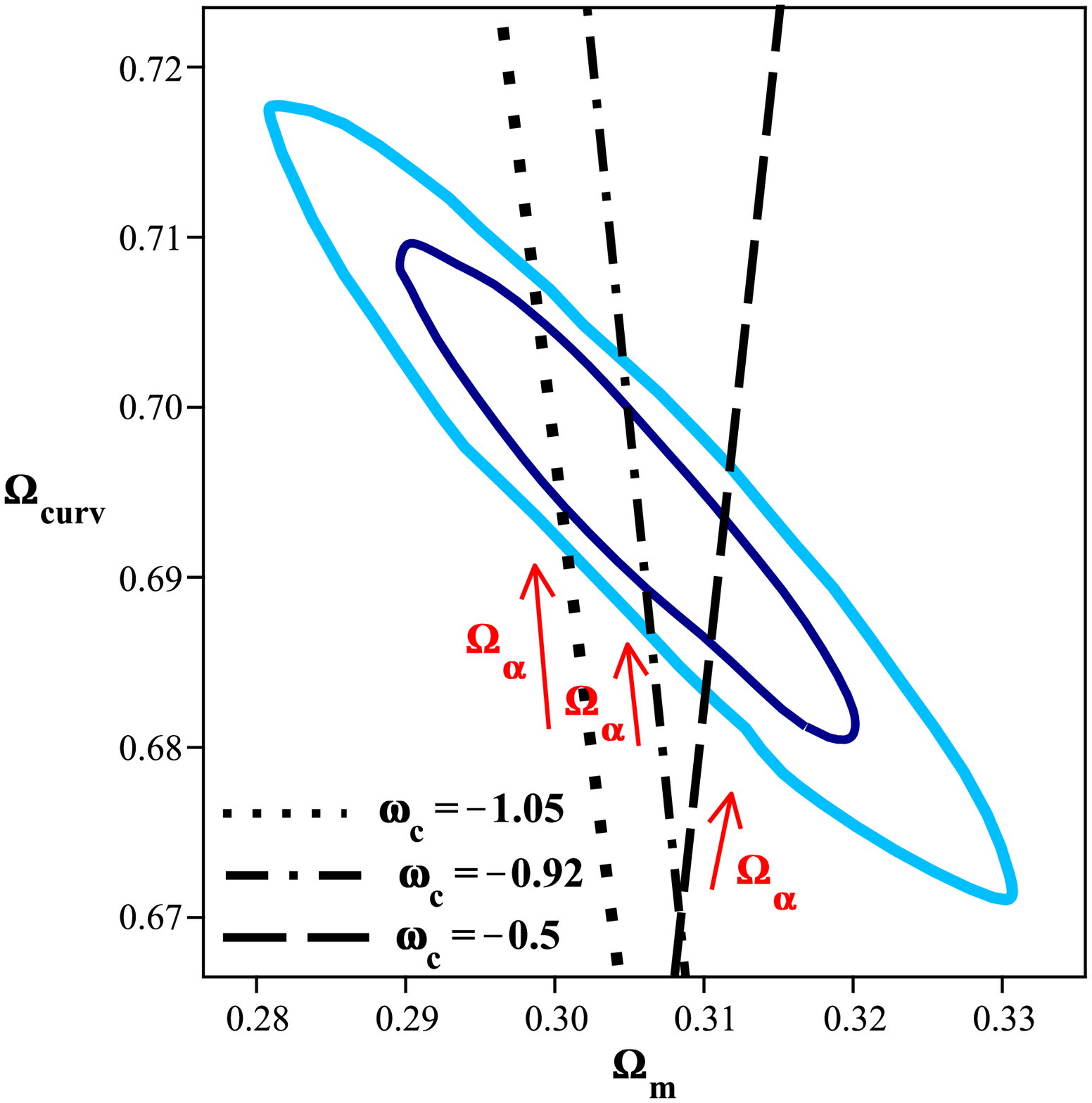} \vspace{6cm}
\end{center}
\caption{\small { $\Omega_{curv}$ versus $\Omega_{m}$ in the
background of Planck + WMAP + high L+ lensing + BAO joint data.  }}
\end{figure}

\begin{figure}[htp]
\begin{center}\includegraphics{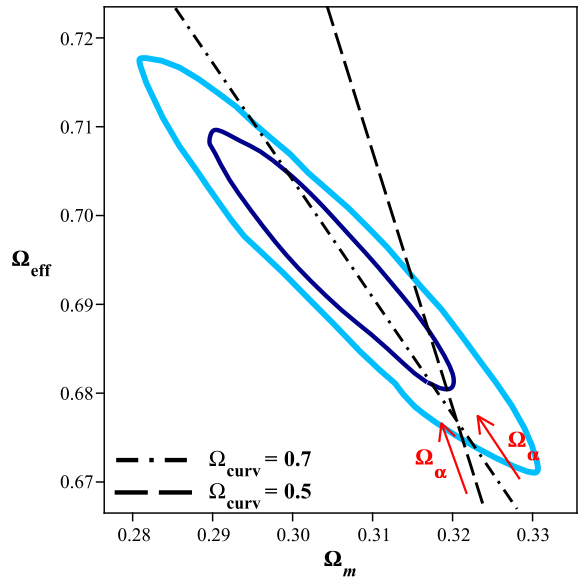} \vspace{6cm}
\end{center}
\caption{\small { $\Omega_{eff}$ versus $\Omega_{m}$ in the
background of Planck + WMAP + high L+ lensing + BAO joint data.  }}
\end{figure}

\section{Summary and Conclusion}
In this paper, we have constructed a DGP-inspired braneworld
scenario where induced gravity on the brane is modified in the
spirit of $f(R)$-gravity and higher order curvature effects are
taken into account by incorporation of the Gauss-Bonnet term in the
bulk action. It is well-known that the normal branch of the DGP
braneworld scenario, which is not self-accelerating, has the
potential to realize phantom-like behavior without introducing any
phantom fields neither on the brane nor in the bulk. Our motivation
here to study this extension of the DGP setup is to explore possible
influences of the curvature corrections, especially the modified
induced gravity, on the cosmological dynamics of the normal branch
of the DGP setup. In this regard, cosmological dynamics of this
scenario as an alternative for dark energy proposal is studied and
the effects of the curvature corrections on the phantom-like
dynamics of the model are investigated. The complete analysis of the
generalized Friedmann equation needs a cosmographic viewpoint to
$f(R)$ gravity, but here we have tried to find an special solution
of this generalized equation via the discriminant function method.
In our framework, effective energy density attributed to the
curvature plays the role of effective dark energy density. In other
words, we defined a \emph{curvature fluid} with varying equation of
state parameter that incorporates in the definition of effective
dark energy density. The equation of state parameter of this
curvature fluid is evolving and the effective dark energy equation
of state parameter has transition from quintessence to the phantom
phase in a non-smooth manner. We have considered a cosmologically
viable (Hu-Sawicki) ansatz for $f(R)$ gravity on the brane to have
more practical results. We have shown that this model mimics the
phantom-like behavior on the normal branch of the scenario in some
subspaces of the model parameter space without introduction of any
phantom matter neither in the bulk nor on the brane. In the same
time, the null energy condition is respected in the phantom-like
phase of the model parameter space. There is no super-acceleration
or big rip singularity in this setup. Incorporation of the curvature
effects both in the bulk (via the Gauss-Bonnet term) and on the
brane (via modified induced gravity) results in the facility that
\emph{curvature fluid} plays the role of dark energy component. On
the other hand, this extension allows the model to mimic the
phantom-like prescription in relatively wider range of redshifts in
comparison to the case that induced gravity is not modified. This
effective phantom-like behavior permits us to study cosmological
dynamics of this setup from a dynamical system viewpoint. This
analysis has been performed with details and its consequences are
explained. The detailed dynamical system analysis of this setup is
more involved relative to the case that there are no curvature
effects. We have shown that with suitable condition on equation of
state parameter of curvature fluid, there is an unstable matter era
and a stable de Sitter phase in this scenario leading to the
conclusion that this model is cosmologically viable. We have
constraint our model based on the recent observational data from
joint Planck + WMAP + high L+ lensing + BAO data sets. In this way
some constraints on Gauss-Bonnet coupling contribution are
presented. We note that no Rip singularity is present in this model
since the Gauss-Bonnet contribution to this model is essentially a
stringy, quantum gravity effects that prevents the Rip singularity
(see for instance [106] for details).\\

{\bf Acknowledgment}\\

The work of K. Nozari has been supported financially by the Center
for Excellence in Astronomy and Astrophysics (CEAAI - RIAAM),
Maragha, Iran.

\end{document}